\documentclass{article} % For LaTeX2e
\usepackage{iclr2025_conference,times}

\usepackage{amsmath,amsfonts,bm}

\def\eqref#1{equation~\ref{#1}}
\def\1{\bm{1}}

\DeclareMathAlphabet{\mathsfit}{\encodingdefault}{\sfdefault}{m}{sl}
\SetMathAlphabet{\mathsfit}{bold}{\encodingdefault}{\sfdefault}{bx}{n}

\usepackage{hyperref}
\usepackage{url}
\usepackage{booktabs}
\usepackage{graphicx}
\usepackage{amsmath}
\usepackage{bm}
\usepackage{xcolor}

\iclrfinalcopy

\title{Privacy-Preserving Text Inference for Health-Misinformation Classification and Semantic Retrieval}

\begin{document}

\author{Jasleen Kaur$^1$, Chloé Cachet$^1$, Koray Karabina$^{2}$\thanks{The work was initiated when the author was affiliated with the National Research Council of Canada.}, Jean Belo Klamti$^1$, Philippe Lamontagne$^1$ \\
$^1$National Research Council Canada, Ottawa, Canada\\
$^{2}$Florida Atlantic University, Boca Raton, USA
}

\maketitle
\begin{abstract}
Health-misinformation analysis presents a dual challenge: misleading claims can affect public well-being, while systems designed to detect them may expose sensitive user-submitted text. This motivates privacy-preserving inference for public-health applications in which user queries may contain confidential medical information and trained models and reference databases are maintained on organizational servers. We design a privacy-preserving text inference framework based on CKKS homomorphic encryption for health-misinformation classification and semantic retrieval. The classification pipeline fuses BioBERT embeddings, PCA-reduced MPNet embeddings, and SVD-reduced TF-IDF features, while the retrieval pipeline combines MPNet embeddings with ISCC-SCT semantic fingerprints through late fusion. We evaluate predictive performance and benchmark plaintext versus ciphertext-based execution. The hybrid classifier achieved an accuracy of 0.971 and a macro-F1 score of 0.950, with identical reported metrics under plaintext and ciphertext-based inference. The hybrid late-fusion similarity model achieved a Pearson correlation of 0.8691 with human sentence-pair similarity scores on STS-B.  For ciphertext-based retrieval, we evaluated naive and optimized implementations. The optimized implementation uses CKKS single instruction, multiple data (SIMD) packing to compute and return multiple similarity scores together, reducing single-query latency over 7,128 indexed items from 222.028 to 5.168 seconds and communication overhead from 6.11 GB to 2.13 MB. These findings demonstrate the technical feasibility of privacy-preserving text inference for health-misinformation applications and characterize its predictive utility and efficiency trade-offs. 
\end{abstract}

\section{Introduction}
\label{sec:introduction}

Artificial intelligence (AI) has transformed how information is created, distributed, and consumed while introducing new challenges related to content authenticity, privacy, and trust. These challenges are especially significant in health, where misleading claims about treatments, vaccines, disease transmission, and prevention can affect public well-being, health-related decision-making, and trust in institutions. The World Health Organization characterizes the overabundance of accurate and inaccurate health information as an ``infodemic,'' which can create confusion, encourage risky behaviour, and complicate public-health responses \cite{WHO2026}. Machine-learning (ML) and natural-language-processing (NLP) approaches have evolved from traditional classifiers based on lexical and statistical features to transformer-based models that capture contextual and semantic relationships\cite{XJYDL25,ZWOFM25}. Combining lexical and semantic representations can provide richer linguistic information and improve misinformation detection \cite{FBHDK24,AZHHBAZJCM25}, but it also increases feature dimensionality and computational cost.

Automated misinformation-detection systems introduce an important privacy trade-off because user-submitted text or its high-dimensional representation is commonly transmitted to a remote inference service. Health-related user queries may contain symptoms, medical histories, treatment details, or other confidential information, and representing them as text embeddings does not eliminate the risk of exposing sensitive information. This motivates privacy-preserving inference for public-health applications in which user queries require protection while trained classification models and indexed reference databases remain on organizational servers. Homomorphic encryption offers a potential solution by enabling computation directly over encrypted data without exposing the underlying plaintext to the processing server \cite{CKS25}. The CKKS homomorphic encryption scheme is particularly suitable for ML inference because it supports approximate arithmetic over real-valued vectors \cite{CKKS17}. Although previous studies have demonstrated privacy-preserving NLP and encrypted text classification \cite{GMHSJ22,FCZ25}, comparatively limited attention has been given to designing and evaluating encrypted text-inference frameworks for health-misinformation applications, including classification and semantic retrieval, while examining the effects of representation choice on predictive performance and system efficiency.

This study develops a privacy-preserving text inference framework for ciphertext-based health misinformation classification and semantic retrieval in a client-server architecture\footnote{Only inference is performed on encrypted data, model training is performed on plaintext data.}. Classification uses feature-level fusion of biomedical, paraphrase-based, and lexical representations, while retrieval combines contextual and fingerprint-based similarity through score-level late fusion. The client generates and encrypts the query representations, which the server processes only in encrypted form. This study makes three contributions: (1) the design of a privacy-preserving text inference framework for ciphertext-based health-misinformation classification and semantic retrieval; (2) the integration and comparison of hybrid text representations using early fusion for classification and score-level late fusion for retrieval; and (3) plaintext versus ciphertext-based benchmarking of predictive performance, computational cost, communication overhead, and the effects of ciphertext packing.

\section{Proposed Privacy-Preserving Framework}
\label{sec:framework}
The proposed framework supports two privacy-preserving text-inference pipelines: health misinformation classification and semantic retrieval. Classification can flag potentially misleading health claims, while retrieval can search an indexed database for semantically related content. Both pipelines share a client-server architecture for health-misinformation applications where user queries may contain sensitive medical information. Text representation and encryption are performed by the client, whereas classification parameters and the indexed database remain on the server. HE enables server-side processing of encrypted query representations without requiring distribution of the model parameters and database to clients. Classification uses feature-level fusion and semantic retrieval uses hybrid score-level late fusion, with both scoring computations adapted for CKKS-compatible arithmetic.

\subsection{Hybrid Representation Fusion}
\label{subsec:hybrid-fusion}

\noindent\textbf{Classification:} The classification pipeline combines three representations that capture complementary linguistic information. Biomedical context is represented using the \textit{S-BioBert-snli-multinli-stsb} sentence-embedding model, which is based on BioBERT \cite{JWSDSHJ20}. Paraphrase-level semantics are captured using \textit{paraphrase-mpnet-base-v2} \cite{KXTJT20}, while TF--IDF features \cite{GC88} represent lexical and term-distribution patterns. To reduce the feature dimensionality and encrypted computational cost, the $768$-dimensional MPNet representation is reduced to $256$ dimensions using Principal Component Analysis (PCA), and the TF--IDF representation is reduced to $300$ dimensions using Truncated Singular Value Decomposition (SVD). The $768$-dimensional BioBERT representation is retained. 

For an input $T$, the client generates the BioBERT representation $x_B=\phi_B(T)$, the reduced MPNet representation $x_M=P^{\top}\phi_M(T)$, and the reduced lexical representation $x_T=U^{\top}\operatorname{TFIDF}(T)$. The learned projections and standardization parameters are applied on the client, and early feature-level fusion concatenates the representations as $x=[x_B \mid x_M \mid x_T]\in\mathbb{R}^{1324}$.

An $L_2$-regularized logistic-regression classifier is trained on the fused vector, with linear decision function $z=x^{\top}w+b$, where $w$ and $b$ denote the learned weights and bias. This function was chosen for its compatibility with the additions and multiplications supported efficiently by CKKS \cite{CKKS17}.

\noindent\textbf{Semantic retrieval:} The retrieval pipeline combines $768$-dimensional MPNet embeddings with $256$-bit ISCC-SCT semantic fingerprints \cite{ISCC25}. MPNet captures contextual similarity, while ISCC-SCT provides a compact fingerprint-based representation. Because these representations have different structures, their similarity scores are computed independently. For a query $q$ and database item $d_i$, hybrid score-level late fusion combines the rescaled MPNet cosine similarity $s_{M,i}$ in $[0,1]$, and normalized ISCC-SCT Hamming similarity $s_{I,i}$ using $L_2$-regularized Ridge regression as $\hat{s}_i=\beta_0+\beta_Ms_{M,i}+\beta_Is_{I,i}$, where $\beta_0$, $\beta_M$, and $\beta_I$ are the effective learned coefficients after including score standardization. This strategy preserves the similarity measure appropriate to each representation while combining the two scores into a learned similarity prediction.

\subsection{Privacy Model and Ciphertext-Based Inference}
\label{subsec:encrypted-inference}

The framework follows a client-server setting with an honest-but-curious server \footnote{The server follows the defined protocol but tries to learn additional information from the received queries.}. The client $C$ holds the input text $T$, the public key $pk$, and the secret key $sk$, whereas the server $S$ holds the trained classification parameters or the indexed retrieval database. Text encoding, feature transformation, encryption, decryption, and final decision-making are performed by the client. Consequently, the server does not receive the plaintext text or its plaintext feature representation. Privacy protection applies during inference, when the server processes encrypted queries using trained classification parameters or indexed database representations; model training is performed on plaintext data.

\noindent\textbf{Ciphertext-based classification:}
The client constructs the fused feature vector $x$ and encrypts it under CKKS as $\widetilde{x}=\operatorname{Enc}(pk,x)$. The encrypted vector $\widetilde{x}$ is sent to the server, which evaluates the logistic-regression score using the plaintext model parameters as $\widetilde{z}=\widetilde{x}^{\top}w+b$.

The server returns $\widetilde{z}$ to the client, which decrypts the score and computes the probability and predicted label:
\begin{equation}
z=\operatorname{Dec}(sk,\widetilde{z}), \qquad
p=\frac{1}{1+e^{-z}}, \qquad
\hat{y}=
\begin{cases}
1, & p\geq 0.5,\\
0, & \text{otherwise}.
\end{cases}
\label{eq:classification-decision}
\end{equation}

The nonlinear sigmoid and threshold operations are performed after decryption to reduce the multiplicative depth and computational cost of the encrypted computation.

\noindent\textbf{Ciphertext-based semantic retrieval: } For retrieval, the client generates and encrypts the query representations before sending them to the server. The query and database MPNet embeddings $e_q$ and $e_i$ are unit-normalized, with similarity computed as $s_{M,i}=(1+e_q^\top e_i)/2$. For binary query and database fingerprints $q_I,d_{I,i}\in\{0,1\}^{L}$, normalized Hamming similarity is $s_{I,i}=\frac{1}{2}+\frac{(2q_I-1)^\top(2d_{I,i}-1)}{2L}$, where $L=256$. The server evaluates the Ridge-fusion score $\widetilde{s}_i=\beta_0+\beta_M\widetilde{s}_{M,i}+\beta_I\widetilde{s}_{I,i}$, for $i=1,\ldots,n$, by combining score standardization and fusion coefficients into a single dot-product calculation with an added constant, using the encrypted query embedding and signed fingerprint against indexed plaintext vectors. The naive implementation computes and returns a separate encrypted score for each database item, whereas the optimized implementation uses CKKS SIMD packing to compute multiple scores together and return a single packed ciphertext in the evaluated configuration. After decryption, the client clips the scores to $[0,1]$ and ranks the database items. Packing reduces both communication overhead and decryption time.

\section{Results}
\label{sec:results}

\subsection{Experimental Setup}
\label{subsec:setup}

The evaluation examines predictive performance and compares plaintext and ciphertext-based execution in terms of computational cost and communication overhead for classification and semantic similarity scoring. Classification was evaluated on the CoAID health-misinformation dataset using a test set of $1{,}081$ samples, comprising $184$ fake and $897$ real samples, with performance measured using accuracy, class-specific F1 scores, and macro-F1. The experiments compared individual and fused combinations of BioBERT, MPNet, and TF-IDF representations. The plaintext and ciphertext-based implementations used the same trained logistic regression parameters to determine whether the approximation introduced by CKKS affected the classification performance. Semantic similarity was evaluated on the Semantic Textual Similarity Benchmark (STS-B) test split using Pearson and Spearman correlations, MSE, and MAE. The score-standardization parameters and Ridge models were learned from the training split. For the fusion model and each single-component calibration baseline, we tested $\alpha \in \{0.0001,0.001,0.01,0.1,1,10,100\}$ and selected the value yielding the lowest MSE on the validation split. All three models selected $\alpha=0.0001$. The selected models were then evaluated on the test split. The calibration baselines distinguish score-calibration effects from fusion benefits. Retrieval latency and communication overhead were benchmarked separately using a single query over $7{,}128$ indexed items. Ciphertext-based inference was implemented using TenSEAL~\cite{TenSEAL2021} with a CKKS polynomial modulus degree $N=2^{15}$, coefficient-modulus bit sizes $[60,40,40,60]$, and a global scale of $2^{40}$. Experiments were conducted on an Apple M4 MacBook Air with 24~GB of unified memory running macOS.

\subsection{Classification and Retrieval Results}
\label{subsec:classification-retrieval-results}

\noindent\textbf{Classification:}
Table~\ref{tab:classification} summarizes the classification ablation study. The full hybrid representation achieved the strongest performance, with an accuracy of $0.971$ and a macro-F1 score of $0.950$. Compared with BioBERT alone, feature-level fusion improved macro-F1 by $0.045$; compared with MPNet alone, the improvement was $0.051$. The fusion combines biomedical context from BioBERT, paraphrase-level semantics from MPNet, and lexical representations from TF-IDF to capture different aspects of health-misinformations claims. However, the configurations differ in dimensionality, so the observed improvement cannot be attributed solely to complementary representations. Plaintext and ciphertext-based inference produced identical accuracy and F1 scores for every feature configuration. Thus, within the reported precision, the CKKS approximation did not alter the classification decisions. The hybrid classifier required approximately $122.4$~ms for encryption, server-side scoring, and decryption, with total encrypted communication $2.133$~MB per query.

\begin{table}[!htbp]
\centering
\caption{Classification performance and ciphertext-based inference cost over $1{,}081$ held-out test samples}
\label{tab:classification}
\resizebox{\columnwidth}{!}{%
\begin{tabular}{lrrrrrr}
\toprule
\textbf{Representation} &
\textbf{Dim.} &
\textbf{Accuracy} &
\textbf{F1-Fake} &
\textbf{F1-Real} &
\textbf{Macro-F1} &
\textbf{Total HE (s)} \\
\midrule
TF-IDF & 300 & 0.942 & 0.844 & 0.964 & 0.904 & 94.347  \\
BioBERT & 768 & 0.944 & 0.844 & 0.966 & 0.905 & 79.918  \\
MPNet  & 256 & 0.938 & 0.837 & 0.962 & 0.899 & 44.230  \\
BioBERT + MPNet & 1,024 & 0.967 & 0.906 & 0.980 & 0.943 & 51.754  \\
BioBERT + TF-IDF & 1,068 & 0.957 & 0.878 & 0.974 & 0.926 & 97.476  \\
\textbf{Full hybrid} & \textbf{1,324} &
\textbf{0.971} & \textbf{0.918} & \textbf{0.983} &
\textbf{0.950} & \textbf{132.789} \\
\bottomrule
\end{tabular}%
}
\end{table}

\noindent\textbf{Semantic retrieval:}
Table~\ref{tab:semantic_similarity} compares sentence-pair similarity scores from MPNet and ISCC-SCT, their individually Ridge-calibrated scores, and late Ridge fusion. Calibration adjusts similarity scores to the human annotation scale. The fusion combines contextual similarity from MPNet with fingerprint-based similarity from ISCC-SCT. It achieved the highest Pearson correlation ($r=0.8691$) and lowest prediction errors ($\mathrm{MSE}=0.0232$ and $\mathrm{MAE}=0.1170$), while MPNet alone achieved the highest Spearman correlation ($\rho=0.8699$, versus $0.8682$ for fusion). Most of the error reduction was obtained through calibration, with fusion providing modest additional gains over calibrated MPNet. These results evaluate sentence-pair similarity rather than retrieval effectiveness. Table~\ref{tab:retrieval_performance} compares plaintext retrieval with naive and optimized ciphertext-based query execution. In both ciphertext-based implementations, the server compares the encrypted query representation with plaintext database representations. In the naive implementation, a separate encrypted similarity result is produced for each database item, whereas the optimized implementation uses CKKS SIMD packing to compute and return multiple similarity scores within a single ciphertext in the evaluated configuration. This optimization reduced single-query latency from $222.028$~s to $5.168$~s and communication overhead from $6.11$~GB to $2.13$~MB per ciphertext-based retrieval query, corresponding to a $43.0\times$ reduction in latency and approximately three orders of magnitude reduction in communication. Plaintext execution required $0.057$~s for the same query.

\begin{table}[!htbp]
\centering
\caption{Semantic similarity performance on the STS-B test split}
\label{tab:semantic_similarity}
\begin{tabular}{lcccc}
\toprule
\textbf{Representation} &
\textbf{Pearson} &
\textbf{Spearman} &
\textbf{MSE} &
\textbf{MAE} \\
\midrule
ISCC-SCT & 0.8302 & 0.8298 & 0.0938 & 0.2445 \\
MPNet & 0.8625 & \textbf{0.8699} & 0.1327 & 0.3084 \\
ISCC-SCT + Ridge calibration &
0.8312 & 0.8298 & 0.0289 & 0.1323 \\
MPNet + Ridge calibration &
0.8666 & 0.8697 & 0.0237 & 0.1192 \\
Late Ridge fusion &
\textbf{0.8691} & 0.8682 &
\textbf{0.0232} & \textbf{0.1170} \\
\bottomrule
\end{tabular}
\end{table}

\begin{table}[!htbp]
\centering
\caption{Single-query retrieval latency and communication over $7{,}128$ database items}
\label{tab:retrieval_performance}
\resizebox{\columnwidth}{!}{%
\begin{tabular}{lrrrrrr}
\toprule
\textbf{Mode} &
\textbf{Encode (s)} &
\textbf{Encrypt (s)} &
\textbf{Server (s)} &
\textbf{Decrypt (s)} &
\textbf{Total (s)} &
\textbf{Communication} \\
\midrule
Plaintext
& 0.053 & 0.000 & 0.004 & 0.000 & 0.057 & 63.69 KB \\
Naive Ciphertext-based
& 0.053 & 0.008 & 210.695 & 11.272 & 222.028 & 6.11 GB \\
\textbf{Opt. Ciphertext-based}
& \textbf{0.053} & \textbf{0.008} & \textbf{5.105} &
\textbf{0.002} & \textbf{5.168} & \textbf{2.13 MB} \\
\bottomrule
\end{tabular}%
}
\end{table}

\section{Discussion \& Conclusions}
\label{sec:discussion-conclusion}

The results demonstrate the technical feasibility of the proposed privacy-preserving text inference framework for health-misinformation classification and semantic retrieval. Hybrid text representations improved performance under the evaluated configurations while remaining compatible with homomorphically encrypted computation. Feature-level fusion combined biomedical context, paraphrase-level semantics, and lexical patterns. This is important because misleading health claims may preserve credible medical terminology while changing the meaning or certainty of the underlying information. Late fusion provided modest gains in Pearson correlation and reductions in prediction error over calibrated MPNet on STS-B. CKKS preserved the reported classification performance without requiring the server to access plaintext query representations, and its SIMD packing substantially reduced ciphertext-based retrieval latency and communication overhead. The framework has been integrated into early-stage public-facing and analytical interfaces for private claim analysis and public-health misinformation exploration, although these interfaces were not formally evaluated in this study. The evaluation remains limited to a pandemic-focused classification dataset and the general-domain STS-B benchmark, with comparisons among component representations and fusion configurations rather than external baselines. External baseline comparisons, dimension-matched classification ablations, and health-specific retrieval evaluation beyond STS-B remain future work. These findings establish a technical foundation for privacy-preserving health-misinformation analysis and clarify its predictive utility and efficiency trade-offs.

\section*{Acknowledgments}
The project was supported by the Canadian AI Safety Institute (CAISI).

\bibliography{iclr2025_conference}
\bibliographystyle{iclr2025_conference}

\end{document}